\documentclass[conference]{IEEEtran}
\IEEEoverridecommandlockouts
\usepackage{cite}
\usepackage{amsmath,amssymb,amsfonts}
\usepackage{algorithmic}
\usepackage{graphicx}
\usepackage{textcomp}
\usepackage{xcolor}
\def\BibTeX{{\rm B\kern-.05em{\sc i\kern-.025em b}\kern-.08em
    T\kern-.1667em\lower.7ex\hbox{E}\kern-.125emX}}

\makeatletter
\newcommand{\linebreakand}{%
  \end{@IEEEauthorhalign}
  \hfill\mbox{}\par
  \mbox{}\hfill\begin{@IEEEauthorhalign}
}
\makeatother

\newcommand{\eqdef}{\triangleq}
\newcommand{\INRc}{\text{INR}_{\text{c}}}
\newcommand{\INRp}{\text{INR}_{\text{p}}}
\newcommand{\SPNR}{\text{SPNR}}
\newcommand{\SNR}{\text{SNR}}
\newcommand{\SSR}{\text{SSR}}
\newcommand{\RFI}{\text{RFI}}
\newcommand{\SIRc}{\text{SIR}_{\text{c}}}
\newcommand{\SIRp}{\text{SIR}_{\text{p}}}

\usepackage{tikz}

\newcommand\copyrighttext{%
  \footnotesize \textcopyright \the\year{} IEEE. Personal use of this material is permitted. Permission from IEEE must be obtained for all other uses, including reprinting/republishing this material for advertising or promotional purposes, collecting new collected works for resale or redistribution to servers or lists, or reuse of any copyrighted component of this work in other works.}

\newcommand\copyrightnotice{%
\begin{tikzpicture}[remember picture,overlay]
\node[anchor=south,yshift=10pt] at (current page.south) {%
\begin{minipage}{\textwidth}
\center \copyrighttext
\end{minipage}};
\end{tikzpicture}%
}

\newcommand\acceptedtext{%
  \footnotesize This article has been accepted for publication in proceedings of the 2024 IEEE 11th International Workshop on Metrology for AeroSpace (MetroAeroSpace), but has not been fully edited. Content may change prior to final publication. \\
Citation information: DOI 10.1109/MetroAeroSpace61015.2024.10591529.}
  
\newcommand\acceptednotice{%
\begin{tikzpicture}[remember picture,overlay]
\node[anchor=north,yshift=-6pt] at (current page.north) {%
\begin{minipage}{\textwidth}
\center \acceptedtext
\end{minipage}};
\end{tikzpicture}%
}

\begin{document}

\title{A real/fast-time simulator for impact assessment\\
of spoofing \& jamming attacks on GNSS receivers
\thanks{This paper falls under the activities carried out for the AURORA
(itAlian Urban aiR mObility technologies \& distRibuted test-fAcility) project;
this project has been funded by the European Space Agency
under the NAVISP (NAVigation Innovation in Support Program) program,
activity code NAVISP-EL3-018.}
}


\author{
\IEEEauthorblockN{
Ivan Iudice\IEEEauthorrefmark{1},
Domenico Pascarella\IEEEauthorrefmark{1},
Gianluca Corraro\IEEEauthorrefmark{2},
and
Giovanni Cuciniello\IEEEauthorrefmark{2}
}
\IEEEauthorblockA{
\IEEEauthorrefmark{1}
Security of Systems and Infrastructures Laboratory,
Reliability \& Security Department\\
}
\IEEEauthorblockA{
\IEEEauthorrefmark{2}
Guidance, Navigation and Control Laboratory,
Reliability \& Security Department\\
}
\IEEEauthorblockA{
Italian Aerospace Research Centre (CIRA), Capua (CE), Italy\\
Email: [i.iudice, d.pascarella, g.corraro, g.cuciniello]@cira.it
}
}

\maketitle
\copyrightnotice
\acceptednotice

\begin{abstract}
In aviation, the impact of threats is becoming increasingly significant,
particularly for global navigation satellite system (GNSS).
Two relevant GNSS threats are represented by jamming and spoofing.
In order to evaluate the technological solutions to counter
GNSS attacks, such attacks should be assessed by means of a proper
GNSS threat simulator.
This work shows the implementation and the testing results of
a GNSS security impact simulator which injects the desired
threat scenarios as a deviations on the GNSS actual measurements.
The proposed simulator can be integrated in both real- and fast-time simulation environments.
The provided results confirm the effectiveness of the simulator,
and include in-flight demonstrations by means of a flight experimental vehicle.
\end{abstract}


\section{Introduction}

A (cyber) threat represents any potential cause of a (cyber) security incident,
where a security incident means an intentional event that might be,
or could lead to, an operational interruption, disruption, loss, emergency, crisis, 
compromising integrity, availability or confidentiality in case of cyber impacts \cite{Le2017}.
In aviation, the impact of threats is becoming increasingly significant.
This is particularly true for \emph{global navigation satellite system} (GNSS) since \cite{Wan2022}:
(i) its importance makes it an attractive target for attackers;
(ii) its openness (at least for civil GNSS) makes it considerably
vulnerable;
(iii) its weak transmitting and receiving power makes it susceptible to interferences,
both natural and malicious.
%
%
Two relevant GNSS threats are represented by jamming and spoofing.
\emph{Jamming} refers to intentional transmission of \emph{radio frequency} (RF) energy
to hinder a navigation service by masking GNSS signals with interference.
%
Instead, \emph{spoofing} is a malicious interference made by a \emph{spoofer} to alter
a GNSS receiver measurement, providing fake position, velocity and time information.

In order to evaluate the technological solutions to counter
GNSS attacks, such attacks should be assessed. 
%
Accordingly, a GNSS \emph{Threat Simulator} (GTS) is an essential tool to:
(i) characterize the impact of intentional attacks exploiting GNSS vulnerabilities;
(ii) verify and validate GNSS intrusion detection and mitigation systems.

This work shows the implementation and the testing results of
a GTS as a GNSS \emph{security impact simulator}, i.e., a simulator reproducing
the desired threat scenarios by injecting representative deviations on the GNSS actual measurements.
The simulator can be integrated in real/fast-time simulation testbeds
for being used in a laboratory and flight environment
in order to reproduce typical GNSS security threats in emulated/real GNSS receivers.

The provided results are referred to both fast-time and real-time simulations.
For the latter, in-flight demonstrations of specific threat scenarios
were performed by means of a flight experimental vehicle.

\subsection{Related work}

Several studies have addressed the characterization
of commercial jammers and their effect on GNSS receivers.
\cite{Bet2000} compares theoretical predictions
with experimental results to examine the effect of Gaussian interference
on code tracking accuracy.
\cite{Tit2003} evaluates the effects of intrasystem and intersystem interference
by analytical techniques based on established theory
and first-order estimates, in terms of the receiver's
effective carrier-to-noise-density ratio.
In \cite{Liu2010}, the authors propose determining
the minimum acceptable degradation of effective carrier-to-noise-density ratio,
considering all receiver processing phases.
\cite{Mit2012} investigates the simulation,
Kalman Filter tracking, and Kalman Filter geolocation
of a chirp–type civilian GPS jammer.
In \cite{Bor2013}, the impact of jammers on GPS and Galileo 
is investigated, showing that the signals of each system are affected
in similar way.

About spoofing, 
\cite{Gei2018} provides a preliminary risk assessment
at system and aircraft level for potential cyberattacks
to the Flight Management System and to GNSS receivers.
\cite{Wan2022} proposes a GNSS-induced spoofing simulation algorithm
based on path planning, which generates spoofing signals
by adjusting the related powers.
\cite{Cou2020} reports a general mathematical model
of a received spoofed GNSS signal, considering the meaconing cyberattack.
\cite{Liu2018} analyses the impact of spoofing
on the error covariance of the Intertial Navigation System (INS),
revealing how the used Kalman filter responds to spoofing attacks.
\cite{Zha2013} reports the architecture of a platform
for the simulation of GNSS threats,
including spoofing and implementing enhanced models
to better predict the occurrence of threats. 

To the best of authors’ knowledge,
our work is the first to propose a joint real and fast-time simulator
for impact assessment of spoofing and jamming attacks on GNSS receivers.

\section{System model and implementation}

The proposed GTS is composed by two modules:
the RFI threats module and the Cyber threats module.
The former is intended for injecting RFI threats
by simulating intentional/unintentional
interferences to the target GNSS receiver (e.g., jamming).
The latter injects cyber attacks by simulating
spoofing attacks to the target GNSS receiver.

\subsection{Jamming model}

Jamming includes GNSS attacks
denying GNSS receivers the access to the information
supplied by one or more GNSS satellites.
Thus, jamming affects the availability feature
of GNSS signals for the receivers \cite{Ade2020}.
%

In the RFI threats module of the proposed GTS,
the interference impact is computed in terms of SNR gain.
Different threat scenarios are simulated according to the nature of the interference,
i.e., continuous or pulsed emissions.
The SNR gain is generated as a composition of three distinct contribution:
(i) continuous RFI (e.g., jamming, intersystem interferences),
(ii) pulsed RFI (e.g., transponders, VHF radios),
(ii) cyber threats (e.g., spoofing).

For the $i$-th satellite operating at the frequency $f$,
the overall SNR gain is computed using the following expression,
\begin{equation}
G_{\RFI}^{(i)}(f) \eqdef
\frac{1-\beta}{1+\INRc(f)+\INRp(f,\beta)+\SPNR_i(f)},
\end{equation}
where $\beta$ represents the so-called \emph{blanker} duty cycle,
\begin{equation}
\INRc(f) \eqdef \frac{\SNR_{\text{max}}(f)}{\SIRc(f)},
\end{equation}
\begin{equation}
\INRp(f,\beta) \eqdef \frac{1}{\beta} \sum_{n=1}^{N_\text{p}},
\frac{\SNR_{\text{max}}(f)}{\SIRp^{(n)}(f)} d_n
\end{equation}
\begin{equation}
\SPNR_i(f) \eqdef \SNR_i(f) \, \SSR_i(f),
\end{equation}
with $\SNR_{\text{max}}(f)$ representing the maximum
signal-to-noise power ratio, i.e.,
$\SNR_{\text{max}}(f) \eqdef \max_i \SNR_i(f)$,
$\SIRc(f)$ standing for the overall signal-to-continuous-interference
power ratio, $\SIRp^{(n)}$ and $d_n$ corresponding to the
signal-to-pulsed-interference power ratio and the duty-cycle,
respectively, of the $n$-th pulsed interference, for all
$n \in \{1,2,\ldots,N_\text{p}\}$, and, finally,
$\SSR_i(f)$ referring to the spoofing-to-signal power ratio
of the $i$-th satellite (see Sec. \ref{sec:spoofing-system-model}).

When the receiver is caught by the spoofing on the $i$-th satellite
$\SNR_i(f)$ and $\SPNR_i(f)$ are swapped.
Continuous/pulsed RFI and cyber threats
follow different enabling path.
Cyber threats enabling signal is provided by the spoofing module
(see Sec. \ref{sec:spoofing-system-model}).
Regarding cyber threats, the simulation model assumes that
the spoofing only affects the SNR of the satellite under attack;
this assumption can be accepted given the
(almost-)orthogonality of the chip sequences.

\subsection{Spoofing model}
\label{sec:spoofing-system-model}

Spoofing is a class of GNSS cyberattacks manipulating
the information presented to GNSS receivers \cite{Sch2016}.
%
%
The only detectable differences between legitimate satellite signals
and spoofed ones may be in discrepancies in timing, signal direction,
strength, Doppler shift, 
and signal to noise ratio.
As an example, the simplest form of spoofer is a \emph{meaconer}
which records authentic GNSS signals and plays back
the recorded signals with a time delay and sufficient transmit power \cite{Ade2020}.

The spoofing impact is computed
by the Cyber threats module of the proposed GTS
in terms of pseudorange-drifts and spoofing-to-signal power ratios.
%
For the computation of the pseudorange-drift,
we extend the model available in \cite{Liu2018},
by evaluating the pseudorange $\rho_{i,r}(t)$
between a satellite $i$ and a receiver $r$ as,
\begin{equation}
\begin{aligned}
\rho_{i,r}\left(t\right) & = c \tau_{i,r}
+ c \left[\left(t+\delta t_{r}\right) - \left(t+\delta t_{i}\right)\right] \\
& = c \tau_{i,r} + c \left(\delta t_{r} - \delta t_{i}\right),
\label{eqspoof_rho1}
\end{aligned}
\end{equation}
where $c$ is the speed of light, $\tau_{i,r}$
is the signal transmission delay between $i$ and $r$,
$\delta t_{r}$ is the receiver clock offset from the GNSS time scale,
$\delta t_{r}$ is the clock offset of the satellite from the GNSS time scale.

\begin{figure*}[t]
\centering
\includegraphics[width=\textwidth]{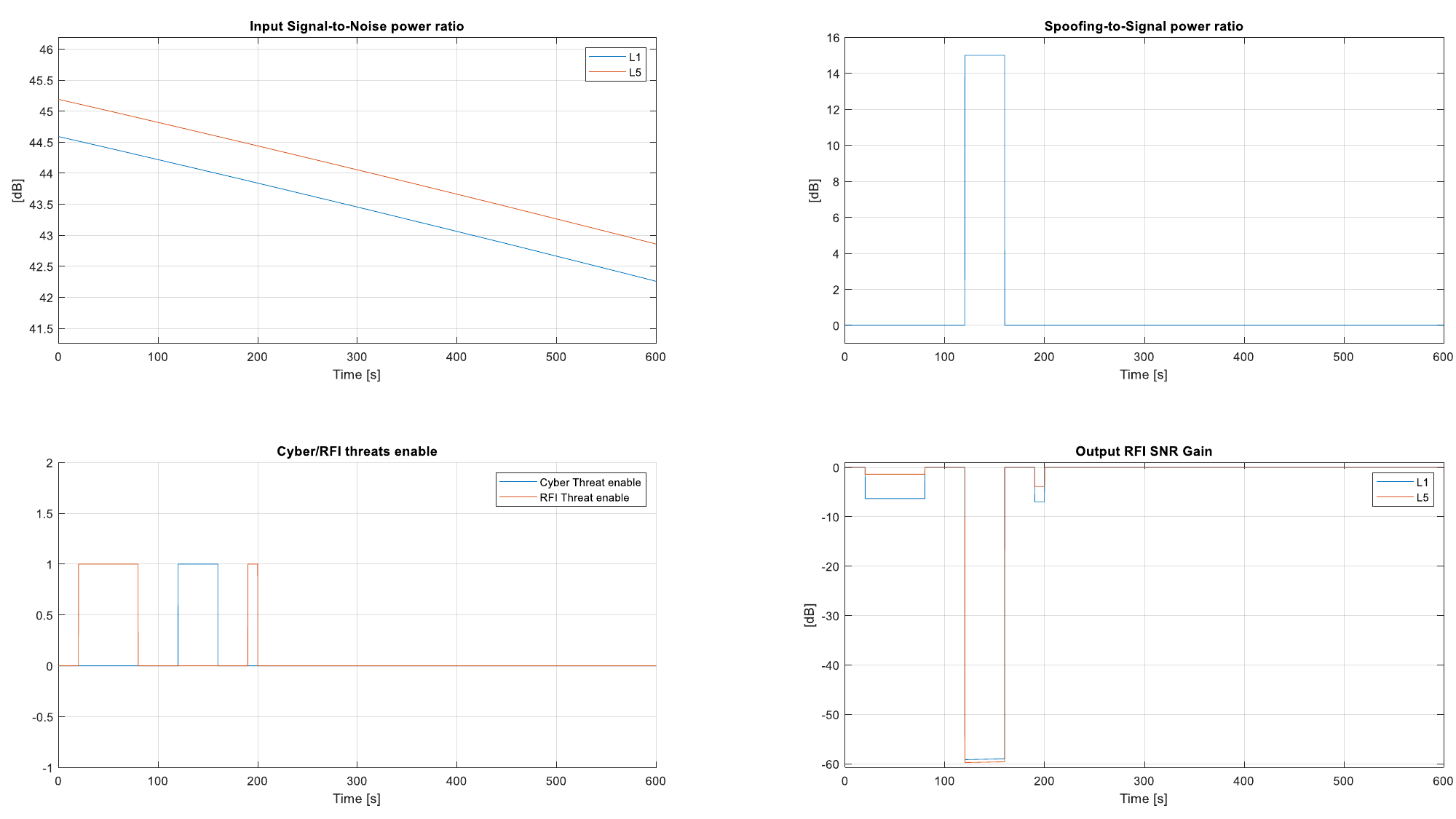}
\caption{RFI/Cyber Threat effects on PRN\#19.}
\label{fig:rfi}
\end{figure*}

For an authentic signal from $i$,
the signal delay $\tau_{i,r}^{(\text{a})}$
may be modelled as,
\begin{equation}
\tau_{i,r}^{(\text{a})} = \frac{d_{i,r}}{c} + I_{i,r} + T_{i,r},
\label{eqspoof_authdelay}
\end{equation}
where $d_{i,r}$, $I_{i,r}$ and $T_{i,r}$ are the geometric range,
the ionospheric delay and the tropospheric delay, respectively,
between $i$ and $r$.
Then, the authentic pseudorange $\rho_{i,r}^{(\text{a})}$
between $i$ and $r$ may be modelled as,
\begin{equation}
\rho_{i,r}^{(\text{a})} = d_{i,r} + c \left(\delta t_{r}
- \delta t_{i}\right) + c \, I_{i,r} + c \, T_{i,r}.
\label{eqspoof_authpseudorange}
\end{equation}

For a spoofed signal related to $i$ and $r$
by means of the spoofer $s$, the signal delay
$\tau_{i,r}^{(\text{s})}$ may be modelled as,
\begin{equation}
\tau_{i,r}^{(\text{s})} = \frac{d_{i,s} + d_{s,r}}{c} + I_{i,s} + T_{i,s}
+ \Delta t^{\left(i\right)}_{s,\text{proc}}
+ \Delta t^{\left(i\right)}_{s,\text{ctrl}},
\label{eqspoof_spoofdelay1}
\end{equation}
where $d_{i,s}$ is the geometric range between
the $i$ and $s$, $d_{s,r}$
is the geometric range between $s$ and $r$,
$I_{i,s}$ is the ionospheric delay between $i$ and $s$,
$T_{i,s}$ is the tropospheric delay between $i$ and $s$,
$\Delta t^{\left(i\right)}_{s,\text{proc}}$ and
$\Delta t^{\left(i\right)}_{s,\text{ctrl}}$ are
the signal-processing delay and 
the signal-controlled delay, respectively,
of $s$ for $i$.

We assume that the atmospheric delays are the same for the receiver and the spoofer, so,
\begin{equation}
\begin{aligned}
\tau_{i,r}^{(s)} & = \frac{d_{i,r}+\left(d_{i,s}+d_{s,r}-d_{i,r}\right)}{c} + I_{i,r} + T_{i,r} \\
& + \Delta t^{\left(i\right)}_{s,\text{proc}} + \Delta t^{\left(i\right)}_{s,\text{ctrl}} \\
& = \left(\frac{d_{i,r}}{c} + I_{i,r} + T_{i,r} \right) \\
& + \left( \frac{d_{i,s} + d_{s,r} - d_{i,r}}{c}
+ \Delta t^{\left(i\right)}_{s,\text{proc}}
+ \Delta t^{\left(i\right)}_{s,\text{ctrl}} \right) \\
& = \tau_{i,r}^{(\text{a})} + \Delta t^{\left(i,r\right)}_{s},
\label{eqspoof_spoofdelay2}
\end{aligned}
\end{equation}
where,
\begin{equation}
\Delta t^{\left(i,r\right)}_{s} \eqdef \frac{d_{i,s} + d_{s,r} - d_{i,r}}{c}
+ \Delta t^{\left(i\right)}_{s,\text{proc}}
+ \Delta t^{\left(i\right)}_{s,\text{ctrl}}.
\label{eqspoof_spoofdelay3}
\end{equation}


The term $\Delta t^{\left(i,r\right)}_{s}$
represents the spoofing signal delay.
In general, we may assume that
$\Delta t^{\left(i\right)}_{s,\text{proc}} = \Delta t_{s,\text{proc}} \). 
Moreover, the spoofer may exhibit a prediction capability of the navigation data bits
of the authentic signal. Thus, this capability introduces a further offset
$\delta t_{s,\text{pred}}$ to be considered in the spoofed pseudorange
$\rho_{i,r}^{(s)}$, which may be modelled as,
\begin{equation}
\begin{aligned}
\rho_{i,r}^{(s)} & = c \, \tau_{i,r}^{(s)} + c \left[
\left(t + \delta t_{r}\right) - \left(t + \delta t_{s,\text{pred}}
+ \delta t_{s}^{(i)} \right) \right] \\
& = c \, \tau_{i,r}^{(\text{a})} + c \, \left(\delta t_{r} - \delta t_{i}\right)
+ c \, \Delta t^{\left(i,r\right)}_{s} - c \, \delta t_{s,\text{pred}} \\
& = \rho_{i,r}^{(\text{a})} + c \, \Delta t^{\left(i,r\right)}_{s}
- c \, \delta t_{s,\text{pred}}.
\label{eqspoof_spoofpseudorange1}
\end{aligned}
\end{equation}


Note that 
%
%
the pseudorange-drift is actually produced if
$\Delta t^{\left(i,r\right)}_{s} - \delta t_{s,\text{pred}}$
is less than the chip period of $i$.
Furthermore, the spoofing model \eqref{eqspoof_spoofpseudorange1}
holds in case the target receiver is single-frequency, instead,
if the target receiver is multi-frequency, the model is extended as,
\begin{equation}
\rho_{i,r}^{(s)} = \frac{1}{1 - \gamma} \, \rho_{i,r}^{(s,\text{L}5)}
- \frac{\gamma}{1 - \gamma} \, \rho_{i,r}^{(s,\text{L}1)},
\label{eqspoof_spoofpseudorange3}
\end{equation}
where $\rho_{i,r}^{(s,\text{L}5)}$ and $\rho_{i,r}^{(s,\text{L}1)}$
are the spoofed pseudoranges for L5 and L1 frequencies, respectively,
according to \eqref{eqspoof_spoofpseudorange1},
and \( \gamma \eqdef \frac{f_{\text{L}1}}{f_{\text{L}5}}^2 \),
with $f_{\text{L}1}$ and $f_{\text{L}5}$ representing
the corresponding carrier frequencies.

Regarding to the evaluation of $\SSR_i(f)$,
non-smart and smart spoofer scenarios are considered.
For the former, the spoofer is simulated by applying a constant signal power.
For the latter, the spoofer is simulated by applying
a time-dependent spoofing power according to a specific control law, e.g.,
the reference signal power may be a ramp between a minimum and a maximum value
in predefined time interval for all the spoofing signals.
When the power reaches the maximum value,
the receiver is locked by the spoofer.
After the lock on a given spoofing signal,
the spoofer keeps the maximum value of power.

\begin{figure*}[t]
\centering
\includegraphics[width=\textwidth]{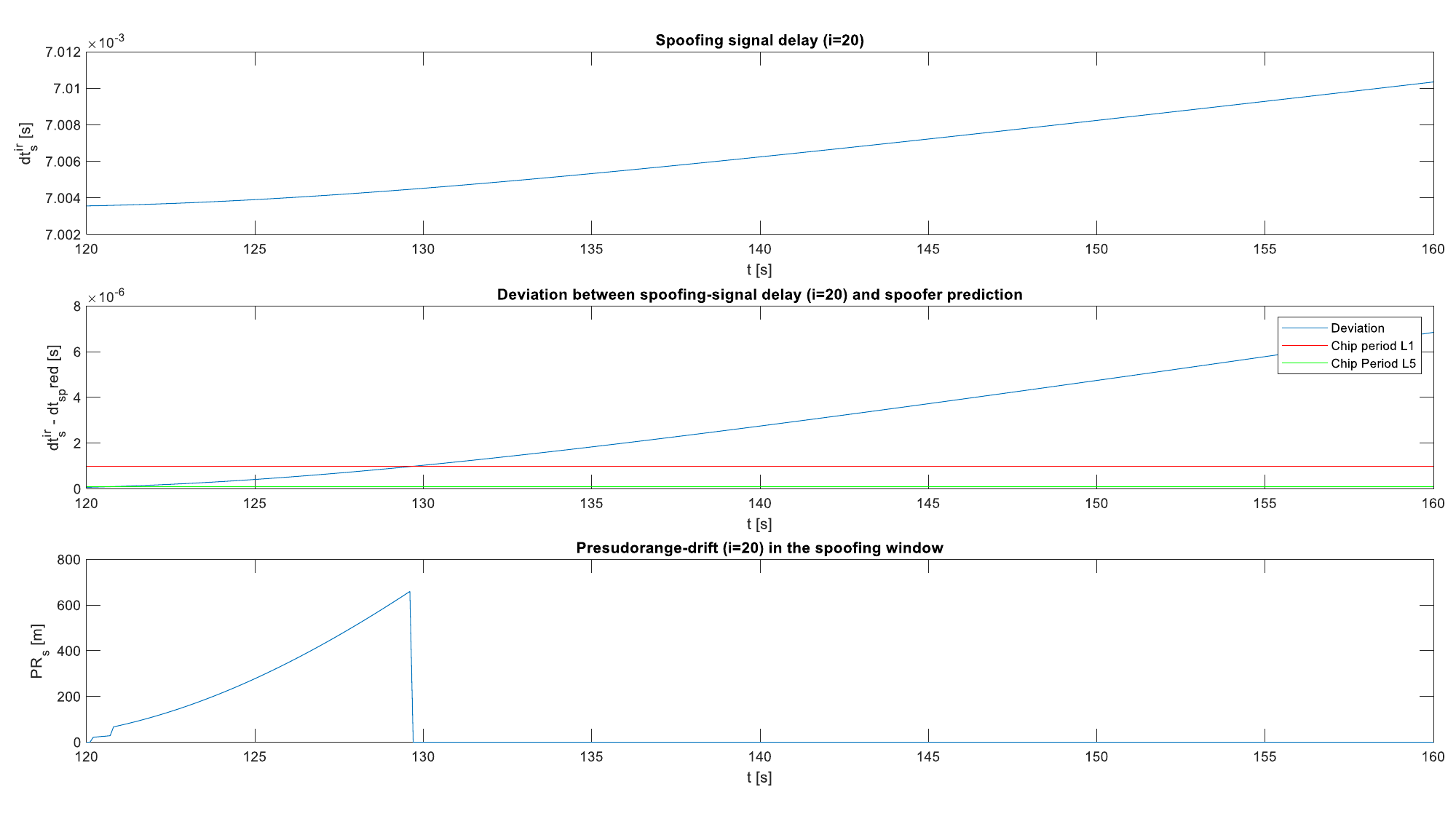}
\caption{Spoofing-signal delay and pseudorange-drifts of PRN\#20.}
\label{fig:spoofing-2}
\end{figure*}

\subsection{GTS implementation}

From a software point of view, the GTS model has been implemented
in MATLAB/Simulink environment.
It modifies the DFMC GNSS measurements in order to introduce
the effects of (cyber) threats as described in the previous section.
Such data are in input together with the \emph{inertial measurement unit} (IMU)
and \emph{air data system} (ADS) measurements to the
\emph{hybrid navigation unit} (HNU) module, that implements all the algorithms
for a tightly coupled GNSS/IMU/ADS sensor fusion with: autonomous
\emph{fault detection and exclusion} (FDE) of GPS/Galileo/BeiDou satellites;
and estimation of navigation solution integrity/accuracy
in terms of \emph{protection level} (PL), and \emph{figures of merit} (FoM).
Finally, the output data are stored in a \emph{flight data recorder} (FDR)
for post-processing analysis.

\section{Test results}

The GTS model was employed to validate a HNU
for UAM applications, by means of both fast-time and \emph{hardware-in-the-loop} (HIL)
real-time simulations, and finally through in-flight demonstration.

\subsection{Fast-time simulation results}

\begin{figure}
\centering
\includegraphics[width=\linewidth]{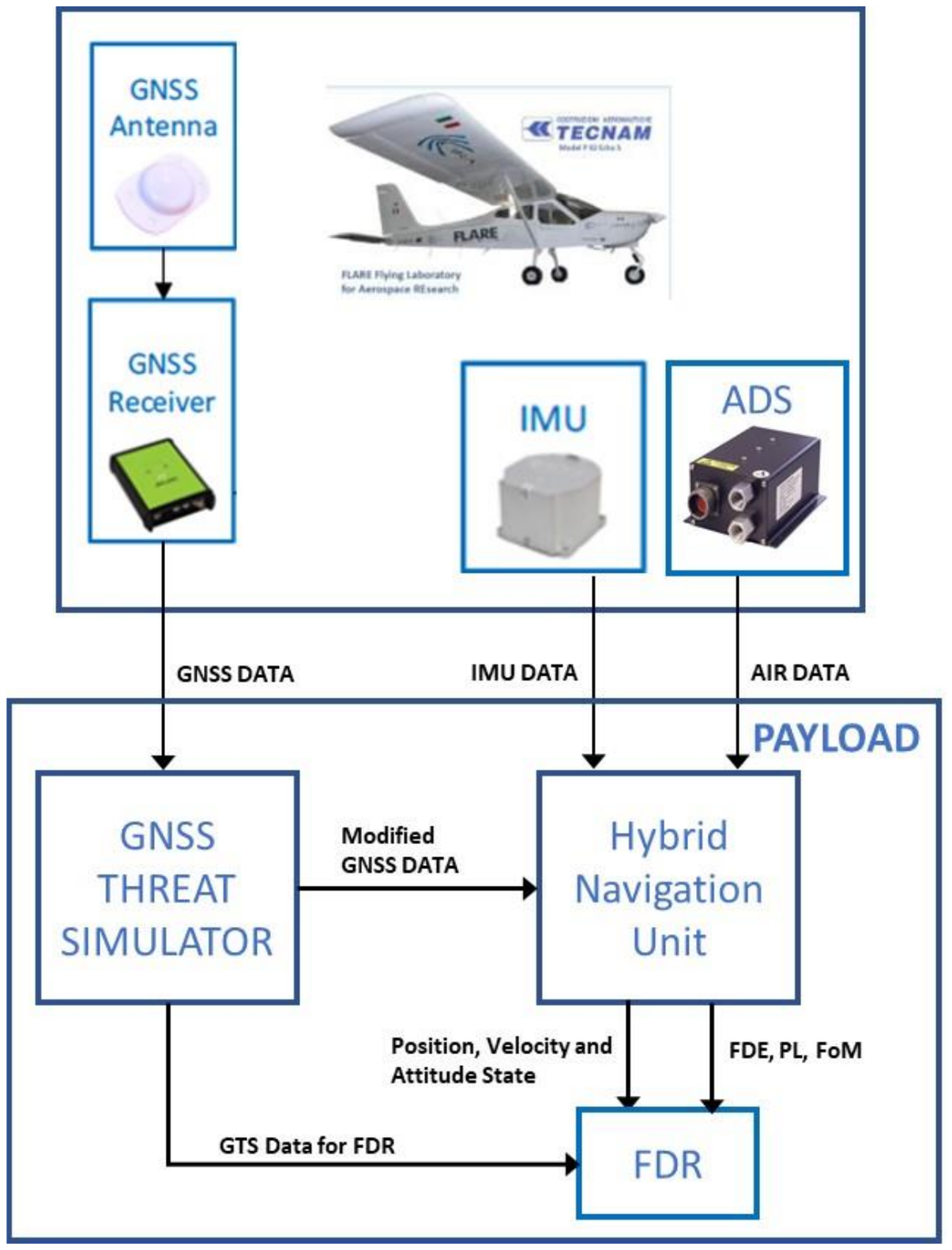}
\caption{In-flight test architecture.}
\label{fig:inflight}
\end{figure}

In the fast-time simulation environment,
the verification of the GTS and HNU modules
has been performed following an architecture
similar to the in-flight one shown in Fig.~\ref{fig:inflight},
implementing simulation models of on-board sensors
and aircraft dynamics.

The fast-time simulation tests were carried-out
on a time span of $10$ min.
Continuous interference was injected
between $20$ and $80$ s,
spoofing was injected between $120$ and $160$ s, finally,
pulsed interference was injected between $190$ and $200$ s.
For the continuous RFI threat was set
a signal-to-interference power ratio of $50$ dB and $60$ dB
for L1 and L5 bands, respectively.
The spoofer was set in non-smart mode,
with a spoofing-to-signal power ratio of $15$ dB.
Three pulsing interferences were considered,
with all of them exhibiting a peak signal-to-interference power ratio
of $40$ dB and $45$ dB in L1 and L5 bands, respectively,
and $3\%$, $4\%$, and $5\%$ as duty-cycle.
The blanker duty-cycle of the receiver was disabled.

%
As you can see in Fig. \ref{fig:rfi}, the first enabling window
corresponds to the continuous interference which produces
negative gains in both L1 and L2 bands.
The second enabling window is related to spoofing;
in this case the huge spoofing-to-signal power ratio
produces a relevant attenuation of the SNR.
The third enabling window consists on the pulsed inteferences;
note that despite the power of pulsed sources quite greater
than the continuous one, the overall effect on the SNR attenuation
is almost the same of the continuous interference due to
the pulsed nature of the interferences.

Regarding the spoofing, it also impacts the pseudorange estimates; thus,
two other simulations were settled for
non-smart and smart spoofer scenarios, both between $120$ and $160$ s.
A static spoofer and a multi-frequency receiver were simulated.
At the beginning of the spoofing window,
the spoofer is $800$ m far from the mobile target receiver.
For the sake of brevity, here we report only the results in terms of pseudorange-drift,
which are the same for both spoofing types, since these only affect
the spoofing-to-signal power ratios and the receiver-lock status.


%
Fig. \ref{fig:spoofing-2} shows:
(i) the spoofing-signal delay;
(ii) the deviation between the spoofing-signal delay and the spoofer prediction,
i.e., $\Delta t_s^{(i,r)} - \delta t_{s,\text{pred}}$;
(iii) the pseudorange-drift.
As you can see, the deviation is always lower than the chip period of L5,
except for an initial time interval; it is lower than the chip period of L1 until 130 s.
As a consequence, the pseudorange-drift exhibits a first high-variation,
i.e., when the deviation becomes greater than the chip period of L5.
Then, the pseudorange-drift is cancelled when the deviation is greater
than the chip period of L1, too, coherently with \eqref{eqspoof_spoofpseudorange3}.

\subsection{Real-time flight results}

The in-flight demonstrations were performed on a flight experimental vehicle
based on a TECNAM P-92 ECHO S, modified in order to perform as a flying testbed
for validation of autonomous flight technologies.

Fig.~\ref{fig:inflight} depicts the architecture of the in-flight setup,
with the following hardware devices:
\begin{itemize}
\item a GNSS receiver, model JAVAD-DELTA3,
and related antenna capable to receive GPS, Galileo and BeiDou signals;
\item an IMU,
model Civitanavi Kryo, based on fiber-optic gyros and quartz accelerometers;
\item an ADS;
\item a PC-104 platform for the real-time execution of the items under test
(i.e., GTS and HNU modules), not involved in the control of the automatic flight
(payload module in Fig.~\ref{fig:inflight}).
\end{itemize}

In the in-flight demonstration campaign,
two scenarios were tested using the GTS
to simulate the impact of spoofing and jamming threats.
Such simulations correspond to a static spoofing/jamming
carried-on from the roof of a building located quite close
to the flight trajectory and powered on when the vehicle
is proximal to the spoofer location,
about $<4.5$ km.

\begin{figure}[t]
\centering
\includegraphics[width=\columnwidth]{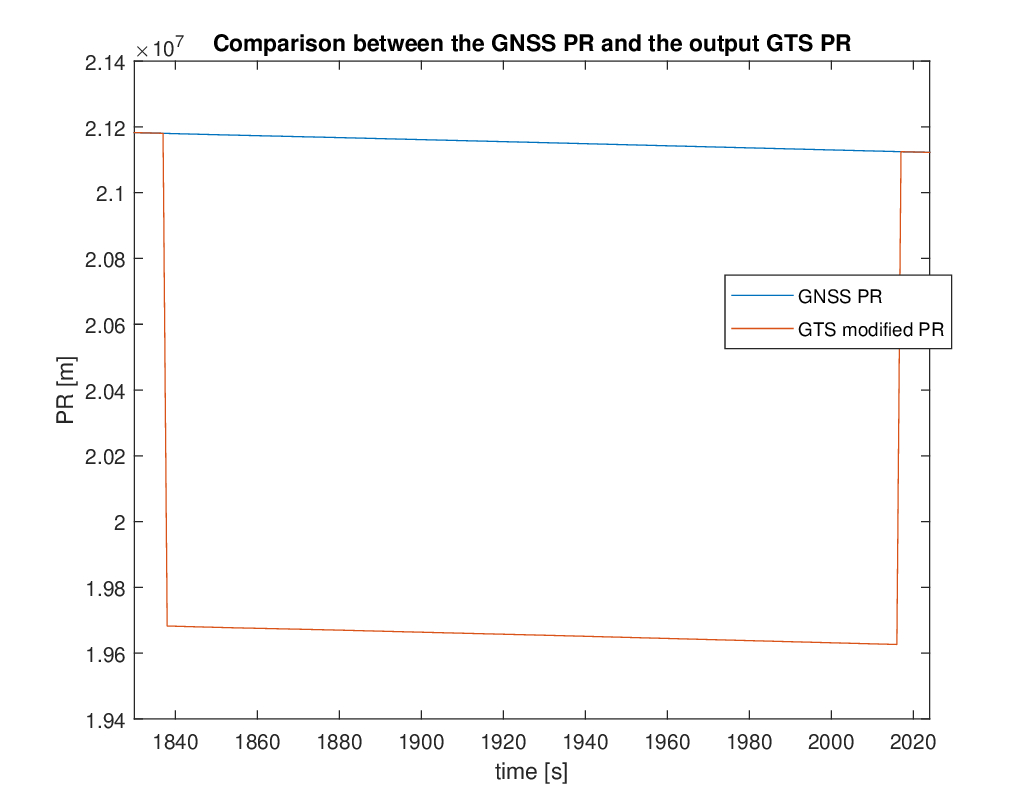}
\caption{Input and spoofed pseudoranges in real-time simulation.}
\label{fig:flight-spoofing-1}
\end{figure}

\begin{figure}[t]
\centering
\includegraphics[width=\columnwidth]{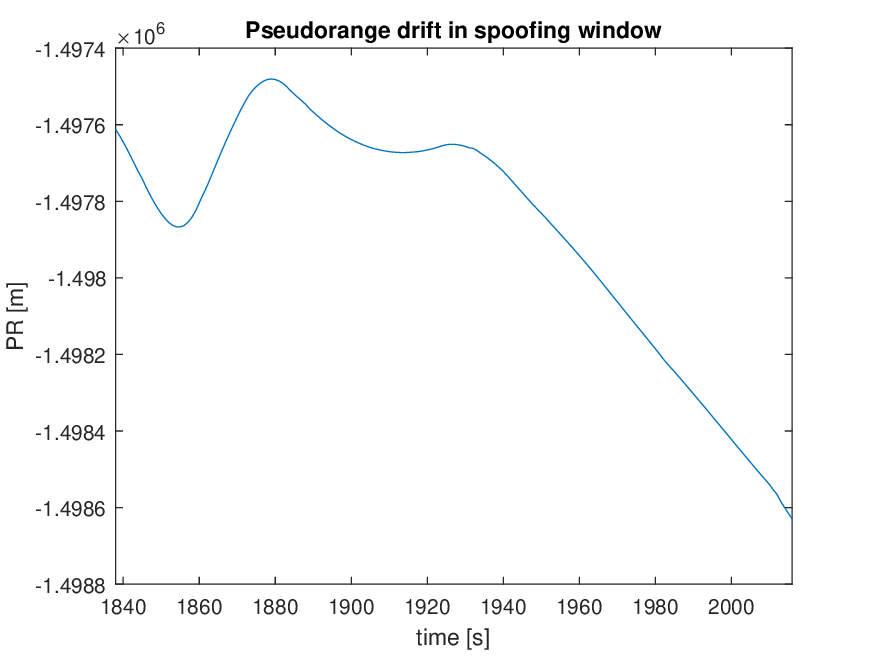}
\caption{Pseudorange drift in real-time simulation.}
\label{fig:flight-spoofing-2}
\end{figure}

For spoofing, the GTS was configured to inject a meaconing
($\delta t_{s,\text{pred}} = 0$) attack
with a spoofing-to-signal power ratio of about $45$ dB,
between $1838$ and $2016$ s.
%
%
%
Fig.~\ref{fig:flight-spoofing-1} shows the comparison between the input
and spoofed pseudorange, whereas Fig.~\ref{fig:flight-spoofing-2} reports
the evaluated pseudorange-drift in the spoofing time window.
The effect of the pseudorange drift changes with the distance
between the spoofer and the receiver.

\begin{figure}[t]
\centering
\includegraphics[width=\columnwidth]{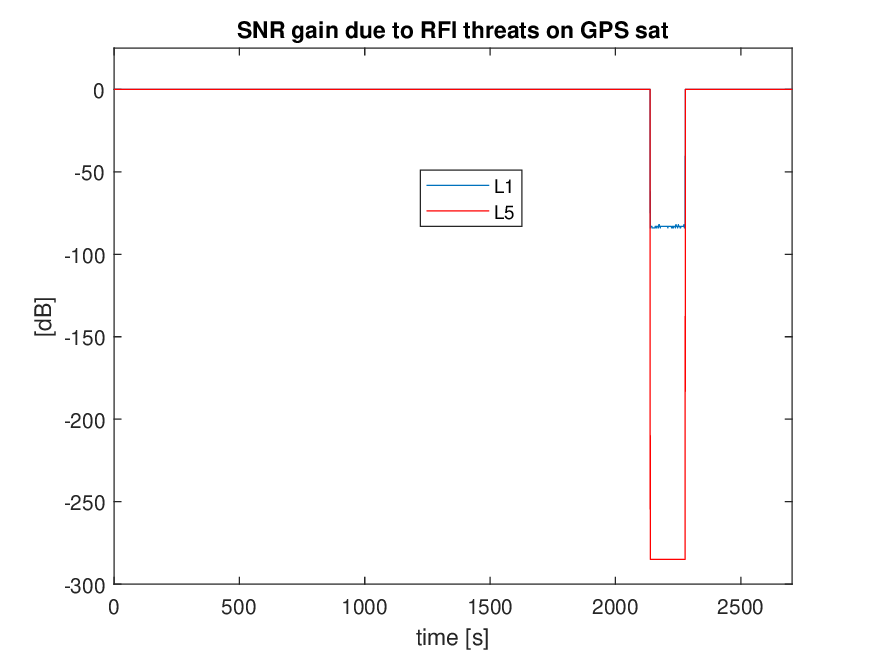}
\caption{SNR gain in real-time simulation.}
\label{fig:flight-jamming}
\end{figure}

For jamming, the GTS was configured
to inject a continuous RFI signal on both L1 and L5 bands,
with a signal power such as to saturate the GNSS receiver RF section,
between $2138$ and $2279$ s.
%
The saturation caused the complete obscuration of satellites,
until the vehicle was out of the effective range of the jammer.
Fig.~\ref{fig:flight-jamming} depicts the effect of the simulated RFI threat
on the SNR gain for one of the GPS satellite in view on both L1 and L5 bands.

\section{Conclusions}

In this paper the implementation and testing results of
a GTS as a GNSS \emph{security impact simulator} are shown.
The simulator can be integrated in real/fast-time simulation environments.
It reproduces RFI and cyber (spoofing) threats in emulated/real GNSS receivers.
The effectiveness of the GTS was proven by fast-time simulations,
as well as by in-flight demonstrations of jamming and spoofing attack scenarios,
carried-out by a flight experimental vehicle.

Future work will provide advanced models for smart spoofing scenarios,
including control laws based on the real trajectory of the GNSS receiver.
Furthermore, the GTS will be extended with lower level models
in order to emulate the physical layer for both jamming and spoofing attacks.

\bibliography{biblio}{}

\begin{thebibliography}{10}
\providecommand{\url}[1]{#1}
\csname url@samestyle\endcsname
\providecommand{\newblock}{\relax}
\providecommand{\bibinfo}[2]{#2}
\providecommand{\BIBentrySTDinterwordspacing}{\spaceskip=0pt\relax}
\providecommand{\BIBentryALTinterwordstretchfactor}{4}
\providecommand{\BIBentryALTinterwordspacing}{\spaceskip=\fontdimen2\font plus
\BIBentryALTinterwordstretchfactor\fontdimen3\font minus
  \fontdimen4\font\relax}
\providecommand{\BIBforeignlanguage}[2]{{%
\expandafter\ifx\csname l@#1\endcsname\relax
\typeout{** WARNING: IEEEtran.bst: No hyphenation pattern has been}%
\typeout{** loaded for the language `#1'. Using the pattern for}%
\typeout{** the default language instead.}%
\else
\language=\csname l@#1\endcsname
\fi
#2}}
\providecommand{\BIBdecl}{\relax}
\BIBdecl

\bibitem{Le2017}
M.~Le~Fevre, B.~G{\"o}lz, R.~Flohr, T.~Stelkens-Kobsch, and T.~S. Verhoogt,
  ``2.0 security risk assessment methodology for {SESAR} 2020; 02.00. 00,''
  \emph{{SESAR} Joint Undertaking: Brussels, Belgium}, vol.~25, 2017.

\bibitem{Wan2022}
\BIBentryALTinterwordspacing
W.~Wang and J.~Wang, ``{GNSS} induced spoofing simulation based on path
  planning,'' \emph{{IET} Radar, Sonar \& Navigation}, vol.~16, no.~1, pp.
  103--112, 2022. [Online]. Available:
  \url{https://ietresearch.onlinelibrary.wiley.com/doi/abs/10.1049/rsn2.12167}
\BIBentrySTDinterwordspacing

\bibitem{Bet2000}
J.~W. Betz, ``Effect of narrowband interference on {GPS} code tracking
  accuracy,'' in \emph{Proceedings of the 2000 National Technical Meeting of
  The Institute of Navigation}, 2000, pp. 16--27.

\bibitem{Tit2003}
B.~M. Titus, J.~Betz, C.~Hegarty, and R.~Owen, ``Intersystem and intrasystem
  interference analysis methodology,'' in \emph{Proceedings of the 16th
  International Technical Meeting of the Satellite Division of The Institute of
  Navigation ({ION GPS/GNSS} 2003)}, 2003, pp. 2061--2069.

\bibitem{Liu2010}
W.~Liu, S.~Li, L.~Liu, M.~Niu, and X.~Zhan, ``A comprehensive methodology for
  assessing radio frequency compatibility for {GPS}, galileo and compass,'' in
  \emph{Proceedings of the 23rd International Technical Meeting of The
  Satellite Division of the Institute of Navigation ({ION GNSS} 2010)}, 2010,
  pp. 943--954.

\bibitem{Mit2012}
R.~H. Mitch, M.~L. Psiaki, B.~W. O'Hanlon, S.~P. Powell, and J.~A. Bhatti,
  ``Civilian {GPS} jammer signal tracking and geolocation,'' in
  \emph{Proceedings of the 25th International Technical Meeting of the
  Satellite Division of The Institute of Navigation ({ION GNSS} 2012)}, 2012,
  pp. 2901--2920.

\bibitem{Bor2013}
D.~Borio, C.~O'Driscoll, and J.~Fortuny, ``Jammer impact on galileo and {GPS}
  receivers,'' in \emph{2013 International Conference on Localization and {GNSS
  (ICL-GNSS)}}.\hskip 1em plus 0.5em minus 0.4em\relax IEEE, 2013, pp. 1--6.

\bibitem{Gei2018}
R.~M. Geister, J.-P. Buch, D.~Niedermeier, G.~Gamba, L.~Canzian, and
  O.~Pozzobon, ``Impact study on cyber threats to {GNSS} and {FMS} systems,''
  2018.

\bibitem{Cou2020}
M.~Coulon, A.~Chabory, A.~Garcia-Pena, J.~Vezinet, C.~Macabiau, P.~Estival,
  P.~Ladoux, and B.~Roturier, ``Characterization of meaconing and its impact on
  {GNSS} receivers,'' in \emph{Proceedings of the 33rd International Technical
  Meeting of the Satellite Division of The Institute of Navigation ({ION GNSS+}
  2020)}, 2020, pp. 3713--3737.

\bibitem{Liu2018}
\BIBentryALTinterwordspacing
Y.~Liu, S.~Li, Q.~Fu, and Z.~Liu, ``Impact assessment of {GNSS} spoofing
  attacks on {INS/GNSS} integrated navigation system,'' \emph{Sensors},
  vol.~18, no.~5, 2018. [Online]. Available:
  \url{https://www.mdpi.com/1424-8220/18/5/1433}
\BIBentrySTDinterwordspacing

\bibitem{Zha2013}
X.~Zhao, X.~Zhan, and K.~Yan, ``{GNSS} vulnerabilities: simulation,
  verification, and mitigation platform design,'' \emph{Geo-spatial Information
  Science}, vol.~16, no.~3, pp. 149--154, 2013.

\bibitem{Ade2020}
\BIBentryALTinterwordspacing
E.~I. Adegoke, M.~S. Bradbury, E.~Kampert, M.~D. Higgins, T.~Watson, P.~A.
  Jennings, C.~R. Ford, G.~Buesnel, and S.~Hickling, ``{PNT} cyber resilience :
  a lab2live observer based approach, report 1 : {GNSS} resilience and
  identified vulnerabilities. technical report 1,'' March 2020. [Online].
  Available: \url{https://wrap.warwick.ac.uk/139519/}
\BIBentrySTDinterwordspacing

\bibitem{Sch2016}
\BIBentryALTinterwordspacing
D.~Schmidt, K.~Radke, S.~Camtepe, E.~Foo, and M.~Ren, ``A survey and analysis
  of the {GNSS} spoofing threat and countermeasures,'' \emph{{ACM} Comput.
  Surv.}, vol.~48, no.~4, may 2016. [Online]. Available:
  \url{https://doi.org/10.1145/2897166}
\BIBentrySTDinterwordspacing

\end{thebibliography}
\bibliographystyle{IEEEtran}

\end{document}